# How to Disclose? Strategic AI Disclosure in Crowdfunding

Ning Wang

School of Business

University of Connecticut

ning.wang@uconn.edu

Chen Liang

School of Business

University of Connecticut

chenliang@uconn.edu

**Abstract**

As artificial intelligence (AI) increasingly integrates into crowdfunding practices, strategic disclosure of AI involvement has become critical. Yet, empirical insights into how different disclosure strategies influence investor decisions remain limited. Drawing on signaling theory and Aristotle's rhetorical framework, we examine how mandatory AI disclosure affects crowdfunding performance and how substantive signals (degree of AI involvement) and rhetorical signals (logos/explicitness, ethos/authenticity, pathos/emotional tone) moderate these effects. Leveraging Kickstarter's mandatory AI disclosure policy as a natural experiment and four supplementary online experiments, we find that mandatory AI disclosure significantly reduces crowdfunding performance: funds raised decline by 39.8% and backer counts by 23.9% for AI-involved projects. However, this adverse effect is systematically moderated by disclosure strategy. Greater AI involvement amplifies the negative effects of AI disclosure, while high authenticity and high explicitness mitigate them. Interestingly, excessive positive emotional tone (a strategy creators might intuitively adopt to counteract AI skepticism) backfires and exacerbates negative outcomes. Supplementary randomized experiments identify two underlying mechanisms: perceived creator competence and AI washing concerns. Substantive signals primarily affect competence judgments, whereas rhetorical signals operate through varied pathways: either mediator alone or both in sequence. These findings provide theoretical and practical insights for entrepreneurs, platforms, and policymakers strategically managing AI transparency in high-stakes investment contexts.

**Keywords:** Artificial Intelligence, Information Disclosure, Crowdfunding, Investment Decisions, Signaling Theory

# 1. INTRODUCTION

The increasing integration of artificial intelligence (AI) into business operations has intensified concerns regarding its potential risks and unintended consequences, prompting regulatory bodies to introduce stricter disclosure requirements. Legislative initiatives such as California's SB 1047[1] and the EU AI Act[2] mandate transparency in AI deployment, aiming to improve oversight and ensure responsible AI use. Concurrently, an emerging stream of research has investigated how AI transparency shapes individual perceptions and decisions, highlighting its effects on trust (Renieris et al., 2024; Schanke et al., 2024), consumer engagement (Carney et al., 2024; Luo et al., 2019), and employee performance (Tong et al., 2021).

These dynamics take on particular importance in crowdfunding. Traditional investment contexts offer informational anchors such as audited financials, third-party verification, and prior relationships to assess quality. Crowdfunding backers lack such anchors, making creator disclosure (Bhargava et al., 2024; Cason et al., 2025; Fu et al., 2025; Kim et al., 2022; Lin & Viswanathan, 2016) the primary channel for assessing creator competence and project quality. When AI is involved, disclosure becomes especially consequential: backers are left to interpret how AI contributes to the project and what it implies for human involvement and capability. This consideration is particularly relevant in the early-stage creative projects typical of crowdfunding. Consequently, different ways of framing AI usage can lead to markedly different interpretations among prospective backers, making crowdfunding a natural setting for examining how disclosure strategies shape economic behavior.

---

[1] California Senate Bill 1047 is available at https://leginfo.legislature.ca.gov/faces/billNavClient.xhtml?bill_id=202320240SB1047 (accessed June 5, 2025).
[2] The European Parliament and Council's Regulation (EU) 2024/1689, known as the EU AI Act, is available at https://www.europarl.europa.eu/topics/en/article/20230601STO93804/eu-ai-act-first-regulation-on-artificial-intelligence (accessed June 5, 2025).

Despite growing regulatory and scholarly attention to AI transparency, existing research focuses primarily on the binary decision of *whether or not* to disclose AI (e.g., Bauer et al., 2025; Carney et al., 2024; Luo et al., 2019; Schanke et al., 2024). Much less is known about the potential role of *AI disclosure strategies* in shaping consequential stakeholder decisions in the digital economy, particularly in settings that involve high-stakes economic commitments and rely heavily on inference, such as crowdfunding. Understanding how specific disclosure strategies encourage or deter investment is, therefore, essential for both theory and practice in the digital economy.

Given that both the content and the style of disclosure may shape backer inferences, we examine two complementary forms of signaling in AI disclosure: substantive signals (Connelly et al., 2011) and rhetorical signals (Suddaby & Greenwood, 2005; Vaara et al., 2016). Substantive signals convey fact-based information about a creator's capabilities or the project development process (Connelly et al., 2011; Sahlman, 1990), offering backers tangible cues about expected product quality (Steigenberger & Wilhelm, 2018). Applied to AI disclosure, substantive signals capture the actual deployment of AI within a project ("AI involvement"), informing backers' judgments about project quality and creator competence.

Rhetorical signals, in contrast, reflect language-based features such as communication style, tone, and presentation (Suddaby & Greenwood, 2005; Vaara et al., 2016) that shape interpretation independent of factual content. Guided by Aristotle's rhetorical triangle (Aristotle, 1991), we focus on three rhetorical dimensions of AI disclosure: explicitness (logos), referring to the level of detail provided; authenticity (ethos), referring to the perceived credibility and trustworthiness; and emotional tone (pathos), referring to the degree of positivity conveyed in the disclosure. While the importance of substantive and rhetorical cues is well-established in contexts such as financial reporting (Feldman et al., 2010), technology licensing (Truong et al., 2022), and public health communication (Hou et al., 2024), it remains unclear how these cues influence pledge decisions in the context of AI disclosure. This gap is consequential: unlike traditional technologies that primarily augment human capabilities, generative AI can complement, substitute for, or constrain human input (Song et al. 2024; Hou et al. 2025; Zhang et al. 2025), while exhibiting far greater opacity and complexity. These features complicate quality assessment and create strategic challenges in both what is disclosed and how, challenges largely absent from prior technology contexts. Motivated by these gaps, we examine:

*1) How does AI disclosure affect crowdfunding success?*

*2) How does this impact differ across different AI disclosure strategies?*

To answer these questions, we collected data from Kickstarter, which introduced its AI disclosure policy in August 2023. The policy requires creators to be transparent and specific about how they use AI in their projects. Following prior literature, we use a keyword-based approach to identify projects that employ AI (Babina et al., 2024; Lou & Wu, 2021; Wu et al., 2025), both before and after the policy, and treat these projects as the treatment group, with all other projects serving as the control group. Leveraging this policy change, we employ a difference-in-differences (DID) design to estimate the causal effect of AI disclosure on crowdfunding performance and examine how the impact of AI disclosure varies with different disclosure strategies.

Our findings reveal that AI disclosure significantly decreases crowdfunding performance. Specifically, the introduction of the AI disclosure policy leads, on average, to a 39.8% decline in funds raised and a 23.9% decrease in the total number of backers for AI-related projects. This negative effect is systematically moderated by both substantive and rhetorical signals. Regarding substantive signals, greater AI involvement amplifies the adverse impact on funding outcomes. Among rhetorical signals, high explicitness (logos) and high authenticity (ethos) mitigate the negative effect, while high positive emotional tone (pathos) exacerbates it.

To uncover the underlying mechanisms driving these moderating effects, we conduct four scenario-based randomized online experiments on Prolific. Each experiment manipulates one disclosure dimension—AI involvement, explicitness, authenticity, or emotional tone—through a between-subjects design where participants evaluate either a high or low version of that dimension. Manipulations are constructed by editing real Kickstarter AI disclosures: one condition presents the original text (edited minimally for readability); the other inverts the focal dimension via Large Language Model (LLM) modification, limiting changes to non-focal content (Bhattacharjee et al., 2024; Youssef et al., 2024). This design allows us to examine how variations in AI disclosure shape investors' perceptions, which in turn influence their willingness to pledge.

Four supplementary online experiments confirm these moderating patterns causally and elucidate the underlying mechanisms. Greater AI involvement decreases pledge intention by reducing perceived creator competence. High explicitness increases pledge intention through two pathways: directly, by enhancing perceived creator competence, and indirectly, by reducing AI washing (i.e., deceptive AI positioning) concerns, which in turn bolster competence perceptions. High authenticity increases pledge intention predominantly through a serial pathway: it first mitigates AI washing concerns, which subsequently enhances perceived creator competence. Conversely, excessively positive emotional tone decreases pledge intention through the same serial pathway but with opposite effects: it first amplifies AI washing concerns, which then diminishes perceived creator competence, thereby lowering pledge intention.

Our study contributes to multiple streams of literature on AI transparency and crowdfunding. First, whereas prior research examines AI disclosure in low-stakes consumption contexts such as customer service (Luo et al., 2019; Xu et al., 2024), content generation (Bauer et al., 2025), and news consumption (Toff & Simon, 2024), where decisions involve minimal resource commitment and are easily reversible, we investigate AI disclosure in settings requiring consequential resource allocation decisions. In crowdfunding contexts, backers must evaluate project quality under significant information asymmetry while committing capital to inherently risky ventures, making disclosure strategies directly consequential for funding outcomes. The diversity of project types and levels of AI involvement provides an ideal setting for examining how AI disclosure shapes investment behavior for early-stage creative ventures.

Second, our study extends the AI transparency literature by shifting the focus from *whether* to disclose AI to *how* to disclose it effectively. As AI transparency becomes increasingly mandated by regulators and platforms, disclosure is no longer a binary choice but an unavoidable requirement. This shift makes the design of disclosure, rather than its mere presence, central to stakeholder interpretation, especially in crowdfunding, where investors rely on narrative cues and inference plays a dominant role. Existing research, however, remains largely focused on the effects of AI disclosure presence versus absence (Baek et al., 2024; Luo et al., 2019; Tong et al., 2021; Xu et al., 2024), offering limited insight into how different disclosure strategies influence perceptions and decision-making once disclosure is required. We address this gap by applying a substantive–rhetorical signaling framework to the AI transparency domain. By distinguishing substantive signals (AI involvement) from rhetorical signals (explicitness, authenticity, emotional tone), we show that disclosure strategy systematically shapes perceived project quality and creator competence, providing a conceptual foundation for understanding AI transparency interpretation when disclosure becomes mandatory.

Finally, our findings contribute to the broader understanding of AI as a unique resource by revealing the underlying mechanisms through which AI disclosure strategies influence decision-making. While prior work documents that AI disclosure affects outcomes, the underlying processes remain largely unexplored, limiting both theoretical understanding and practical guidance. Given AI's unique characteristics (its capacity to offload or substitute human input, combined with inherent complexity and opacity) (Benbya et al., 2021; Lu and Zhang, 2025), we propose that AI disclosure operates through two distinct mechanisms: perceived creator competence and AI washing concerns. Critically, we demonstrate that substantive and rhetorical signals activate these mechanisms differently. Our experimental evidence confirms that substantive signals affect pledge intentions primarily through competence judgments, whereas rhetorical signals exhibit more varied pathways, operating through either mechanism independently or through both in sequence. Our insights advance the theoretical understanding of AI transparency as a unique signaling process and provide actionable guidance for creators, platforms, and policymakers in designing effective disclosure strategies in AI-integrated digital markets.

## 2. LITERATURE REVIEW

### 2.1. AI Transparency in the Digital Economy

The expanding presence of AI in digital services has intensified interest in how organizations communicate algorithmic involvement to users. AI disclosure—broadly defined as revealing when and how AI contributes to a product, service, or decision—has become a central topic in both regulatory debates and academic research (Luo et al. 2019; El Ali et al. 2024; Xu et al. 2024; Wittenberg et al. 2025). Recent policy frameworks, including the EU AI Act and platform-level transparency guidelines issued by firms such as Google[3] and TikTok,[4] reflect a broader movement toward making algorithmic processes more visible, understandable, and accountable to the public. Correspondingly, researchers have begun to document how the disclosure of an AI nature shapes people's trust (Yin et al., 2024), evaluation processes (Chiarella et al., 2022), user engagement (Carney et al., 2024; Xu et al., 2024), and behavioral responses (Tong et al., 2021) across a range of environments. For example, in organizational contexts, disclosing that performance feedback is generated by AI tends to lower employees' trust in the feedback and heighten their concerns about being replaced, which subsequently reduces their learning and job performance (Tong et al., 2021). In the creative industries, the disclosure of AI authorship reduces aesthetic appreciation of artworks (Chiarella et al., 2022). In the logistics industry, identity disclosure of AI voice chatbots reduces response likelihood, whereas anthropomorphic cues (e.g., filler words) positively influence user engagement (Xu et al., 2024).

Meanwhile, emerging work suggests that AI disclosure's consequences are not uniformly adverse. On short-form video platforms, for example, disclosure can increase users' willingness to engage, particularly when the AI is perceived as capable (Chen et al., 2025). Voice-based agents equipped with deepfake voice cloning likewise elicit higher trust, and this trust persists even when AI involvement is disclosed (Schanke et al., 2024). These findings indicate that the impact of AI disclosure in digital contexts is heterogeneous and contingent on the platform, task type, and users' prior expectations.

---

[3] Google's report on ongoing work on responsible AI is available at https://blog.google/technology/ai-responsible-ai-2024-report-ongoing-work/ (accessed November 1, 2025).
[4] TikTok's AI-Generated Content Support Guidelines are available https://support.tiktok.com/en/using-tiktok/creating-videos/ai-generated-content (accessed January 16, 2026).

Existing AI disclosure research conceptualizes disclosure as a binary treatment, i.e., disclosed versus not disclosed, designed to isolate the average effect of an "AI label" on users' trust or engagement. This is highly informative about whether disclosure matters, but it leaves underspecified the managerial question that arises in the digital economy: *how to disclose*. Moreover, much of prior literature focuses on attitudinal reactions or relatively low-stakes engagement metrics (e.g., conversational responses), whereas many consequential decisions in the digital economy involve high-stakes economic commitments. Crowdfunding exemplifies such settings: backers make irreversible investment decisions under information asymmetry, and transparency, while aiding evaluation, may introduce unintended costs (Yang et al., 2022). To fill these gaps, we study how different AI disclosure strategies shape backers' actual investment decisions on Kickstarter. Moving beyond the binary "AI label," we analyze variation in the content and style of disclosure language and test how these disclosure choices translate into downstream campaign funding outcomes.

## 2.2. Signaling in Crowdfunding Platforms

Crowdfunding exemplifies a market characterized by pronounced information asymmetries: creators must persuade a dispersed set of backers who possess limited verifiable information about project feasibility or creator capability (Ahlers et al., 2015; Yang et al. 2016; Courtney et al., 2017; Sabzehzar et al. 2023). Given these asymmetries, prospective backers rely heavily on the content presented in crowdfunding campaigns to assess project quality and make investment decisions. This reliance on such observable signals makes crowdfunding a particularly well-suited context for applying signaling theory to digital markets, where verifiability is limited and informational noise is high (Connelly et al., 2011; Spence, 1973).

Early work emphasizes substantive signals, that is, relatively costly and fact-based indicators that credibly reduce information asymmetry (Bergh et al., 2014; Connelly et al., 2011; Sahlman, 1990). In crowdfunding, prototypes, prior achievements, technical details, and external certification serve as important markers of feasibility and creator/firm competence (Ahlers et al., 2015; Courtney et al., 2017; Xiao et al., 2021). Additional substantive signals that speak to potential sources of uncertainty, such as disclosures of project risk (Kim et al., 2022), unaudited financial statements (Donovan, 2021), additional budget information (Fu et al., 2025), prefunding communication (Wei et al., 2021), timely updates or FAQs (Xiao et al., 2021), also provide backers with relevant information about the likelihood of successful delivery. These signals help backers assess whether creators possess the ability and commitment needed to deliver.

Yet subsequent research shows that substantive cues alone are often insufficient in this high-noise environment, where backers must form impressions quickly and rarely engage in any formal vetting process (Anglin et al., 2025). In such settings, rhetorical strategies also tend to play an important role in conveying relevant information (Anglin et al., 2025; Chandler et al., 2024; Steigenberger & Wilhelm, 2018; Vaara & Monin, 2010). For instance, the rhetorical framing of social responsibility exhibits an inverted U-shaped relationship with funding performance—moderate emphasis proves more effective than either minimal or excessive rhetoric (Anglin et al., 2025). Similarly, rhetorical displays of passion through linguistic markers positively affect social-media exposure and crowdfunding success (Li et al., 2017). At times, these rhetorical signals may strengthen or weaken the influence of substantive signals on a firm's funding performance (Steigenberger & Wilhelm, 2018).

Despite substantial progress in understanding crowdfunding signaling mechanisms, two critical gaps limit our ability to explain strategic disclosure in technology-intensive ventures. First, prior research largely treats technology adoption as an unambiguously positive quality signal, implicitly assuming that more advanced technologies uniformly enhance perceived effort and product quality (Donovan, 2021). This assumption is increasingly untenable in the context of AI, where disclosure could simultaneously signal innovation while raising concerns about reduced creator effort and other potential drawbacks. Moreover, rhetorical strategies effective for traditional technologies may backfire in AI contexts. Unlike earlier technologies that primarily augmented human capabilities, AI can substitute for or constrain human creative labor (Zhang et al. 2024, 2025) while exhibiting high levels of complexity and opacity (Benbya et al., 2021). These characteristics produce ambivalent quality inferences with few precedents in prior technology contexts. We address these gaps by investigating how substantive signals (e.g., the extent of AI usage) and rhetorical signals (i.e., how AI usage is disclosed) jointly shape crowdfunding performance. In doing so, we extend signaling theory to the AI deployment context, where signals are inherently value-laden or contested, and provide actionable guidance for managing AI transparency in entrepreneurial contexts.

## 3. RESEARCH FRAMEWORK

Building on prior work on AI transparency in the digital economy and signaling in crowdfunding platforms, we propose a framework for understanding how creators strategically disclose AI usage. Our framework distinguishes between two complementary dimensions of disclosure: substantive signals, which indicate the extent of AI deployment, and rhetorical signals, which capture how AI deployment is communicated.

### 3.1. Substantive Signals

Substantive signals provide information about the underlying production process, including the resources, capabilities, and effort contributing to an output. In crowdfunding, such signals typically include prototypes, prior achievements, technical specifications, or implementation plans that help reduce information asymmetry by indicating feasibility and competence.

In AI-enabled projects, substantive signals concern *the extent of AI involvement*. These details specify whether AI supports ideation, content generation, refinement, customer interaction, or fulfillment, and whether it augments or substitutes for human creative labor. These distinctions matter: greater AI integration can simultaneously expand production capabilities (Song et al., 2024) while potentially raising concerns about diminished human effort or heightened execution risk (Chiarella et al., 2022; Tong et al., 2021). Substantive signals, therefore, may shape backers' beliefs about creator competence and project quality, ultimately influencing funding decisions.

### 3.2. Rhetorical Signals

While substantive signals communicate what is being offered and how it is produced, rhetorical signals shape how that information is framed and communicated. Rhetorical signals operate through language choice, framing, narrative structure, and communication style, influencing how audiences evaluate and weigh uncertain information (Parhankangas & Renko, 2017). A useful organizing lens is Aristotle's rhetorical triangle, namely, *logos*, *ethos*, and *pathos*, which emphasizes that persuasion depends not only on factual claims but also on rhetorical appeals and affective resonance with audience values (Steigenberger & Wilhelm, 2018). Consistent with this view, policy communication research shows that rhetorical appeals shape responses to political and policy framing (Gottweis, 2017; Stucki & Sager, 2018), and corporate communication research documents similar dynamics in sustainability reporting (Higgins & Walker, 2012).

***Logos-based signals****.* Logos-based communication emphasizes clear, logical reasoning and rational argumentation through specific claims, technical explanations, and factual details (Gottweis, 2017). In crowdfunding settings, signals grounded in clarity and specificity reduce perceived execution risk and increase perceived feasibility (Parhankangas & Renko, 2017). Prior research further shows that narrative structure and language choices that enhance clarity and explicitness shape how backers evaluate project feasibility and, in turn, influence funding outcomes (Moradi et al., 2024). Consistent with this view, risk disclosures that provide relevant details and maintain a balanced tone are more persuasive for projects characterized by high uncertainty (Kim et al., 2022).

To operationalize logos-based rhetorical signals in the context of AI disclosure, we focus on *perceived explicitness*. In this setting, perceived explicitness captures the extent to which a project's AI disclosure provides a clear, logical, and well-reasoned explanation of how AI is used in the project. High perceived explicitness is characterized by language that is specific, informative, and technically descriptive, whereas low perceived explicitness relies on vague, abstract, or lacking in detail.

***Ethos-based signals.*** Ethos-based signals emerge from communication that appears genuine, personally invested, and transparently honest rather than strategically manipulated (Higgins & Walker, 2012). Such signals are often conveyed through authentic narratives that incorporate personal language describing creators' own journeys, candid discussions of challenges, and realistic assessments of project risks. Prior research shows that ethos-based cues, including perceived founder authenticity, enhance backers' trust and emotional warmth toward a project (Radoynovska & King, 2019). Similarly, brand prominence can operate as an ethos-based signal by conveying credibility and reliability (Moradi & Badrinarayanan, 2021).

To capture ethos-based rhetorical signaling in AI disclosure contexts, we focus on *perceived authenticity*. Perceived authenticity refers to the extent to which the disclosure reflects the creator's credibility and trustworthiness through self-revealing and personal expression rather than impersonal and detached communication.

***Pathos-based signals.*** Pathos-based signals refer to the affective valence embedded in campaign communication, which can range from restrained, low-arousal presentations to highly enthusiastic and emotionally charged expressions (Munyon & Summers, 2024). Prior research shows that positive emotional tone, reflected in highly optimistic language and enthusiastic expressions, can generate psychological engagement and identification among audiences (Li et al., 2017). In crowdfunding contexts, pathos-based signals, including emotional appeals embedded in narratives, influence backer attention and engagement (Xiang et al., 2019). Relatedly, evidence from donation and persuasion settings suggests that heightened emotional cues, particularly when they appear exaggerated or incongruent, can intensify affective processing and shape contribution behavior (Yazdani et al., 2025).

In AI disclosure contexts, however, emotional framing may play a more complex role, as overly enthusiastic expressions can amplify both excitement and skepticism toward the technology. To capture this dimension of pathos, we focus on *perceived excessive positive emotion*, measuring the extent to which the tone of AI disclosure conveys exaggerated optimism, inflated enthusiasm, or forced confidence regarding the use of AI.

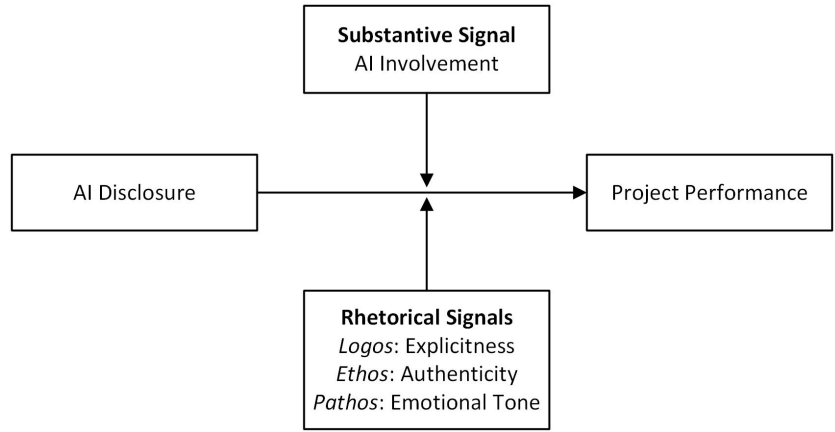

**Figure 1. Research Framework**

Overall, we propose that AI disclosure, which contains both substantive and rhetorical elements, naturally fits within a multidimensional signaling framework. Substantively, disclosure communicates the extent and nature of AI involvement. Rhetorically, creators can frame this involvement through varying levels of explicitness associated with logos-based appeals, authenticity cues reflecting ethos, and emotional tone reflecting pathos. Figure 1 illustrates the proposed research framework.

## 4. EMPIRICAL SETTING AND MODEL

### 4.1. Research Context

Our study draws on data from Kickstarter, one of the largest global crowdfunding platforms, which hosts thousands of campaigns annually across technology, design, and creative categories. On August 29, 2023, Kickstarter introduced a platform-wide disclosure policy requiring creators to declare whether AI tools were used in the development, design, or promotion of their projects.[5] Specifically, during project submission, creators were prompted to complete an AI disclosure section detailing whether and how generative AI contributed to any part of their campaign, such as copywriting, visual design, coding, or prototyping. These disclosures are displayed publicly on each campaign page in a dedicated "Use of AI" section.

---

[5] Kickstarter's disclosure policy is available at https://updates.kickstarter.com/introducing-our-new-ai-policy/ (accessed June 5, 2025).

Our observation window spans one year before and after the policy, covering projects that ended between August 1, 2022, and August 31, 2024. We classify projects into pre-policy and post-policy periods based on their campaign end month rather than launch month. Because the average project duration exceeds one month, projects launched before August 2023 but closing afterward may have been exposed to the policy. Classifying by launch month would misclassify such projects as untreated. Classifying by closing month ensures accurate treatment assignment and yields more conservative estimates.[6]

### 4.2. Keyword-Based Identification of AI-Related Campaigns

Following prior work that exploits policy-induced variation to identify causal effects (e.g., Chen et al., 2011; Chen et al., 2017; Dewan et al., 2017; Huang et al., 2017), we treat the platform's introduction of disclosure requirements as an exogenous shock that applies only to AI-related projects. This policy change generates differential exposure between *AI-related (treated)* and *non-AI-related (control)* projects, thereby enabling difference-in-differences (DID) estimation.

A key empirical challenge is accurately identifying which campaigns are AI-related before and after the implementation of the disclosure requirement. To address this, we follow the existing literature (Babina et al., 2024; Lou & Wu, 2021; Wu et al., 2025) and identify AI-related projects using a keyword-based approach, which has been widely used in categorizing technological concepts and significantly mitigates subjective classification concerns. We develop a comprehensive keyword dictionary to identify crowdfunding campaigns that potentially involve AI usage, which are subject to the disclosure requirement after the policy change but would not have been required to do so had they been posted before the shock.

---

[6] As a robustness check, we classify projects into pre-policy and post-policy periods based on campaign launch date (rather than end date), excluding campaigns that span the policy implementation date. Results are highly consistent (see section 7.2).

To ensure both comprehensiveness and construct validity, we develop a three-part keyword dictionary that captures how AI is described in both technical and practical terms. The first part consists of a foundational set of canonical AI terms (n = 75) from the existing literature (Babina et al., 2024; Miric et al., 2023), including widely recognized terms such as machine learning, deep learning, natural language processing, neural network, and computer vision. These keywords capture the core technical domains consistently referenced in AI research and policy frameworks.

The second part captures how creators describe AI in practice. We extended the dictionary with expressions extracted from campaign descriptions. Specifically, we used GPT-4o-mini to identify AI-related keywords based on the specific questions in AI disclosures (see Online Appendix B1[7] for the prompt). We then parsed all campaigns in our sample and conducted frequency analysis on keywords from campaigns that self-identified as AI-related. We applied three filters: first, manual screening by the authors removed conceptually irrelevant terms; second, we retained only keywords appearing at least three times, which accounted for 83% of all keyword occurrences; third, we excluded terms already present in the canonical AI dictionary from Part 1. This process yielded 27 additional keywords.

The third part reflects the growing prominence and rapid development of generative AI. We supplemented the dictionary with names of major generative AI models, such as ChatGPT, Claude, Gemini, LLaMA, and Mistral (n = 30). Incorporating these terms allows us to capture campaigns that explicitly reference contemporary generative-AI tools.

---

[7] Due to space limitations, the online appendices are hosted on the Open Science Framework (OSF) at https://osf.io/6bcva/overview?view_only=b378f0d57ec9412186292b41c8ea3d8c.

A campaign is classified as AI-related if its title or description contains at least one keyword from this comprehensive codebook. This procedure provides a transparent and reproducible method for AI-related campaigns for our main DID analysis. Importantly, this keyword-based classification captures approximately 97% (1,185 out of 1,220) of campaigns that explicitly disclose AI usage, indicating a high degree of overlap between our operational definition and self-reported AI adoption. All keywords curated during the process are listed in Appendix A.

### 4.3. Variables

*Dependent Variables.* Following the existing literature (Fu et al., 2025; Geva et al., 2024), we use *LogTotalPledge* and *LogTotalBackers* as our main dependent variables. *LogTotalPledge,* is the natural logarithm of the total amount pledged by backers (in U.S. dollars), which captures the overall funding performance of a campaign. In addition, we use *LogTotalBackers*, defined as the natural logarithm of the total number of backers who supported the project, as an alternative dependent variable to measure the reach of campaign engagement.

*Focal Variables.* The key explanatory variable is *Treatment*, a dummy variable that equals 1 if a project is classified as an AI-related project. To account for the timing of the platform policy change, we include a time dummy variable *After*, which equals 1 if the project was closed following the implementation of the AI disclosure policy. Additionally, we include *AIDisclosure*, a dummy variable that equals 1 if a project discloses the use of AI during the post-policy period.

*Moderators.* Drawing on the substantive and rhetorical signals framework, we construct four moderating variables to capture disclosure variation: one substantive dimension (AI involvement) and three rhetorical dimensions (explicitness, authenticity, and emotional tone). We employ GPT-4o-mini to classify these features.[8] Recent research demonstrates that LLMs can effectively perform textual analysis tasks with accuracy comparable to expert human coders while offering substantially greater scalability and cost efficiency (Bail, 2024; de Kok, 2025; Gilardi et al., 2023). LLMs are particularly well-suited for complex classification tasks requiring contextual interpretation and nuanced judgment (de Kok, 2025).

---

[8] We set the temperature parameter to zero to ensure focused and deterministic classifications (de Kok, 2025).

We operationalize the four dimensions as follows: 1) *HighAIInvolvement* is a dummy variable equal to 1 if the project demonstrates extensive AI involvement, where AI was central to producing the project's primary output rather than functioning in a peripheral supporting role (Eloundou et al., 2024), as classified by GPT-4o-mini based on AI disclosures. 2) *HighExplicitness* is a binary variable equal to 1 if the disclosure exhibits high explicitness (above the sample median), where the disclosure provides a clear, logical, and technically descriptive explanation of AI usage rather than vague or general statements (Higgins & Walker, 2012), as scored by GPT-4o-mini. 3) *HighAuthenticity* is a binary indicator equal to 1 if the disclosure demonstrates high authenticity (above the sample median), meaning that the creator conveys credibility and trustworthiness through personal, genuine, and honest writing rather than generic or impersonal language (Bolinger et al., 2024; Higgins & Walker, 2012), as scored by GPT-4o-mini. 4) *HighPosEmotion* is a dummy variable equal to 1 if the disclosure exhibits a highly positive emotional tone, defined as having a positivity score above the sample median, indicating excessively positive emotional tone. This emotion measure is constructed using the Valence Aware Dictionary for Sentiment Reasoning (VADER) package (Hu et al., 2021; Hutto & Gilbert, 2014). All classifications are operationalized via text-based assessment of AI disclosures. Definitions and summary statistics of all aforementioned variables are reported in Table 1.

**Table 1. Definitions and Summary Statistics of Variables**

| Variable | Variable definitions | Obs | Mean | SD | Min | Max |
|---|---|---|---|---|---|---|
| *Dependent variables* | | | | | | |
| *LogTotalPledge* | Total amount (in U.S. dollars) raised by the project (log-transformed) | 35,832 | 7.438 | 2.773 | 0 | 15.714 |
| *LogTotalBackers* | Total number of backers who supported the project (log-transformed) | 35,832 | 3.608 | 1.726 | 0 | 10.515 |
| *Focal variables* | | | | | | |
| *Treatment* | A dummy variable that equals 1 if a project is classified as AI-related | 35,832 | 0.149 | 0.356 | 0 | 1 |
| *After* | A dummy variable that equals 1 if the AI disclosure policy was already introduced in a given month | 35,832 | 0.573 | 0.495 | 0 | 1 |
| *AIDisclosure* | A dummy variable that equals 1 if a project discloses the use of AI during the post-policy period | 20,514 | 0.059 | 0.237 | 0 | 1 |
| *Moderators* | | | | | | |
| *HighAIInvolvement* | A dummy variable equal to 1 if the project demonstrates extensive AI involvement, where AI was central to producing most of the project's output, as classified by GPT-4o-mini based on AI disclosures. | 20,514 | 0.018 | 0.133 | 0 | 1 |
| *HighExplicitness* | A dummy variable equal to 1 if the project demonstrates high disclosure explicitness (i.e., above-median explicitness), where the disclosure provides a clear, logical, and technically descriptive explanation of AI usage rather than vague or general statements, as scored by GPT-4o-mini based on AI disclosures. | 20,514 | 0.032 | 0.176 | 0 | 1 |
| *HighAuthenticity* | A dummy variable equal to 1 if the project demonstrates high disclosure authenticity (i.e., above-median authenticity), where the disclosure reflects the creator's credibility and trustworthiness through personal and honest writing rather than generic or impersonal language, as scored by GPT-4o-mini based on AI disclosures. | 20,514 | 0.034 | 0.180 | 0 | 1 |
| *HighPosEmotion* | A dummy variable equal to 1 if the project demonstrates a high positive emotional tone (i.e., above-median positive sentiment score), as measured by VADER sentiment analysis based on AI disclosures. | 20,514 | 0.030 | 0.171 | 0 | 1 |

Notes: We add 1 to variables before taking the log transformation. The focal variable *AIDisclosure* and four disclosure-related moderators are only observed in post-policy periods. The full sample comprises 35,832 projects across the entire observation period. The post-treatment subsample contains 20,514 projects.

### 4.4. Difference-in-Differences Model

Our DID model is specified in Equation (1). In this model, *Y* represents crowdfunding outcomes, including *LogTotalPledge* (total funding raised) and *LogTotalBackers* (total backers). *After* is a binary variable that equals one if campaigns end following the policy implementation date.[9] We also control for project-specific and creator-specific characteristics (denoted as *X*).

---

[9] Because we define the *After* dummy based on the campaign end date, we can still identify its main effect even after controlling for launch-month fixed effects. As a robustness check, we also re-estimate the model based on

These controls include the length of the project story and title, a dummy variable indicating whether the project story includes any video, the funding goal (in U.S. dollars), the campaign duration, campaign currency dummies to account for systematic differences associated with its original settlement currencies, and the creator's number of previously successful projects, along with category fixed effects (FEs), launch month FEs, and day-of-week FEs. We cluster standard errors at the project category level.

$$Y = \alpha + \beta_1 Treatment + \beta_2 After + \beta_3 Treatment \times After + \boldsymbol{\delta X} + Category\ FE + Launch\ Month\ FE + Day of Week\ FE + \varepsilon. \quad (1)$$

Apart from estimating the effect of the platform's disclosure policy, we also directly measure the effect of AI disclosure itself by comparing projects that include an AI-disclosure statement, captured by the *AIDisclosure* dummy (equals 1 for projects with AI disclosure), to those without disclosure in the post-policy sample. Furthermore, to assess how the treatment effect of AI disclosure varies with the substantive or rhetorical signals of a project (denoted by *Signal*), we incorporate the interaction term $AIDisclosure \times Signal$. To ensure that our estimated effects are not confounded by systematic differences in the types of AI applications disclosed, we further account for potential topic-level heterogeneity in AI usage by introducing the five most common topics identified using GPT-4o-mini as control variables (denoted as **Z**) (Brynjolfsson et al., 2025).[10] The remaining control variables are identical to those used in

---

campaign launch date rather than end date to construct the *After* dummy by excluding campaigns that overlap with the policy implementation period. The results remain consistent with our main findings (see section 7.2).

[10] Here we follow prior work emphasizing the need to account for semantic heterogeneity in text-based measures (e.g., Brynjolfsson et al., 2025). To determine the thematic dimension of AI involvement, we implement a three-step procedure using GPT-4o-mini. First, we feed each AI disclosure to the model and request a short topic phrase (1–3 words) that best summarizes the primary purpose or function of AI in the project. Second, we aggregate all phrases and prompt the model to cluster them into five semantically distinct groups, each with a concise label, a one sentence functional definition, and representative example phrases from the corpus. This procedure yields five categories: Automation and Optimization, Creative Generation, Data Management and Analysis, Support and Assistance, and User Interaction and Personalization. Third, we classify each disclosure into the five categories by asking the model whether the disclosure belongs to any of them (multi label assignment permitted). We then include the resulting topic indicators as additional controls in our regressions to isolate disclosure style effects from

Equation (1). The full specification is presented in Equation (2).

$$Y = \alpha + \beta_1 AIDisclosure + \beta_2 AIDisclosure \times Signal + \boldsymbol{\delta X} + \boldsymbol{\theta Z} + Category\ FE + Launch\ Month\ FE + DayofWeek\ FE + \varepsilon. \quad (2)$$

## 5. EMPIRICAL RESULTS

### 5.1. Parallel Trend Test

A key assumption for the DID model is the parallel trend assumption (Abadie, 2005), which posits that in the absence of treatment, the treatment and control groups would have exhibited similar outcome trajectories. We test this assumption by interacting the treatment group dummy with the month dummies in the following model:

$$Y = \alpha + \sum_{k=-12}^{-2} \eta_k \times Treatment \times Month_{T+k} + \sum_{k=0}^{12} \eta_k \times Treatment \times Month_{T+k} + \boldsymbol{\delta X} + Category\ FE + Launch\ Month\ FE + DayofWeek\ FE + \varepsilon. \quad (3)$$

Here, $Y$ denotes the crowdfunding outcomes, and $T$ represents the month when the AI disclosure policy was introduced by the platform (i.e., August 2023). $\eta_k$ indicates the difference between the treatment and control groups in month $T + k$. The last pre-treatment month ($k = -1$) is set as the baseline. Figure 2 presents the estimated coefficients with 95% confidence intervals, using *LogTotalPledge* and *LogTotalBackers* as the dependent variables, respectively. The estimated lead coefficients are statistically indistinguishable from zero, indicating that the treatment and control groups followed similar trends prior to the platform's implementation of the AI disclosure policy. This provides valid support for our DID design.

---

differences in the substantive nature of AI applications. Full prompts, examples, and coding rules are provided in Online Appendix B3.

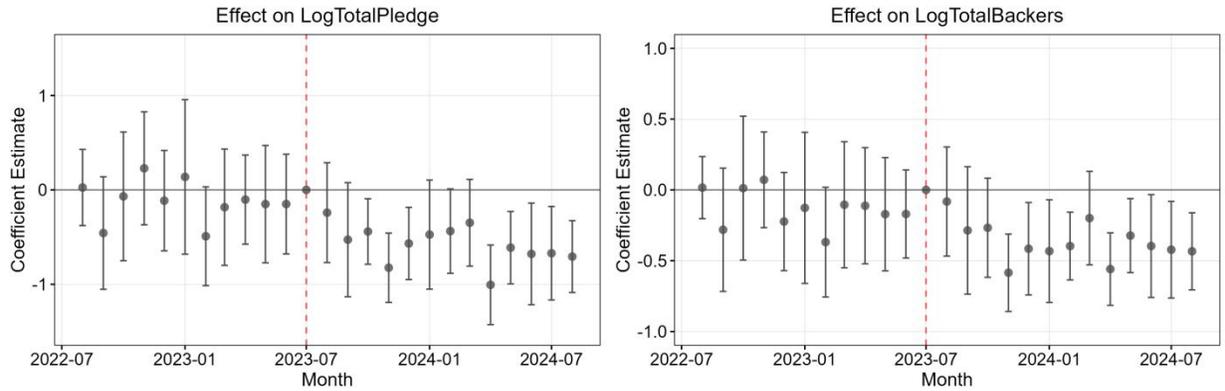

**Figure 2. Parallel Trend Test**

Following the policy change, we observe a significant divergence between these two groups, with AI-related projects experiencing a notable decline in campaign engagement over time. This suggests that AI disclosure may influence prospective backers' evaluations of projects, thereby affecting crowdfunding outcomes. In the next subsection, we will formally quantify these effects using our DID framework.

### 5.2. Average Treatment Effect

Table 2 shows the average treatment effect of AI disclosure on crowdfunding success. Following the implementation of the mandatory AI disclosure policy, projects that disclose AI involvement experience an approximate 39.8% decline in funds raised (calculated as exp(-0.507)-1), and the total number of backers decreases by 23.9% (calculated as exp(-0.273)-1). These results suggest that mandatory AI disclosure significantly reduces backers' participation and funding, posing substantial economic challenges and negatively impacting entrepreneurship. The findings point to the broader economic consequences of AI disclosure mandates and highlight the importance of refining disclosure strategies to alleviate adverse market responses.

## Table 2. Impact of AI Disclosure on Crowdfunding Outcomes

|  | Dependent variable: | |
|---|---|---|
|  | LogTotalPledge | LogTotalBackers |
|  | (1) | (2) |
| Treatment | 0.221*** | 0.079** |
|  | (0.068) | (0.033) |
| After | -0.484** | -0.297*** |
|  | (0.168) | (0.084) |
| Treatment × After | -0.507*** | -0.273*** |
|  | (0.090) | (0.039) |
| Control Variables | YES | YES |
| Category FE | YES | YES |
| Launch Month FE | YES | YES |
| Day-of-week FE | YES | YES |
| Observations | 35,832 | 35,832 |
| R-squared | 0.249 | 0.305 |

Notes: a) We control for project and creator characteristics, including title and description length, funding currency, funding goal, campaign duration, use of video, and the creator's number of prior successful projects. The detailed coefficients for these controls are omitted from the table for brevity. b) Because we define the *After* dummy based on the campaign end date, we can still identify its main effect even after controlling for launch-month fixed effects. The results remain highly consistent when we re-estimate the model using launch date, rather than end date, to construct the *After* dummy. c) Robust standard errors clustered at the category level. * $p<0.1$, ** $p<0.05$, *** $p<0.01$.

### 5.3. Heterogeneous Treatment Effects

We next investigate how the impact of AI disclosure on crowdfunding performance varies across projects that adopt different disclosure strategies. To this end, we restrict our analysis to observations from the post-policy period and compare crowdfunding performance across projects with differing disclosure approaches relative to those without AI disclosure. Guided by our research framework, we consider four potential moderating variables that capture the substantive or rhetorical signals conveyed through projects' AI disclosure statements: *HighAIInvolvement*, *HighExplicitness*, *HighAuthenticity*, and *HighPosEmotion*. This approach allows us to identify heterogeneous effects of AI disclosure across different disclosure strategies and to uncover which forms of AI communication are more effective in attracting backer support.[11]

---

[11] As a robustness check, we control for textual characteristics of project descriptions to address potential confounding. The results remain highly consistent (see Section 7.4).

**Table 3. Moderating Effects of Substantive and Rhetorical Signals in AI Disclosure**

| | Dependent variable: | | | | |
|---|---|---|---|---|---|
| | LogTotalPledge | | | | |
| | (1) | (2) | (3) | (4) | (5) |
| AIDisclosure | -1.349*** | -1.864*** | -1.984*** | -1.227*** | -1.716*** |
| | (0.183) | (0.225) | (0.129) | (0.149) | (0.121) |
| AIDisclosure × HighAIInvolvement | -0.794*** | | | | -0.791*** |
| | (0.255) | | | | (0.199) |
| AIDisclosure × HighAuthenticity | | 0.979*** | | | 0.774** |
| | | (0.206) | | | (0.322) |
| AIDisclosure × HighExplicitness | | | 1.201*** | | 0.715** |
| | | | (0.197) | | (0.273) |
| AIDisclosure × HighPosEmotion | | | | -0.614*** | -0.508*** |
| | | | | (0.140) | (0.163) |
| Control Variables | YES | YES | YES | YES | YES |
| AI Topic Controls | YES | YES | YES | YES | YES |
| Category FE | YES | YES | YES | YES | YES |
| Launch Month FE | YES | YES | YES | YES | YES |
| Day-of-week FE | YES | YES | YES | YES | YES |
| Observations | 20,514 | 20,514 | 20,514 | 20,514 | 20,514 |
| R-squared | 0.265 | 0.266 | 0.267 | 0.265 | 0.269 |

**Table 3 (continued)**

| | Dependent variable: | | | | |
|---|---|---|---|---|---|
| | LogTotalBackers | | | | |
| | (1) | (2) | (3) | (4) | (5) |
| AIDisclosure | -0.816*** | -1.106*** | -1.194*** | -0.736*** | -1.027*** |
| | (0.092) | (0.109) | (0.095) | (0.091) | (0.080) |
| AIDisclosure × HighAIInvolvement | -0.449*** | | | | -0.433*** |
| | (0.150) | | | | (0.133) |
| AIDisclosure × HighExplicitness | | 0.548*** | | | 0.405** |
| | | (0.120) | | | (0.149) |
| AIDisclosure × HighAuthenticity | | | 0.723*** | | 0.463*** |
| | | | (0.113) | | (0.101) |
| AIDisclosure × HighPosEmotion | | | | -0.375*** | -0.313*** |
| | | | | (0.071) | (0.074) |
| Control Variables | YES | YES | YES | YES | YES |
| AI Topic Controls | YES | YES | YES | YES | YES |
| Category FE | YES | YES | YES | YES | YES |
| Launch Month FE | YES | YES | YES | YES | YES |
| Day-of-week FE | YES | YES | YES | YES | YES |
| Observations | 20,514 | 20,514 | 20,514 | 20,514 | 20,514 |
| R-squared | 0.326 | 0.327 | 0.328 | 0.326 | 0.329 |

Notes: a) We include the same project and creator controls as Table 2. Coefficients are omitted for brevity. b) In addition to the control variables included in the previous DID model, we also account for topic-level heterogeneity in AI usage by incorporating dummy variables for the five most common topics extracted from the AI-disclosure statements. These topics capture the primary purpose or application of AI described in each disclosure: Automation and Optimization, Creative Generation, Data Management and Analysis, Support and Assistance, and User Interaction and Personalization. Including these topic indicators ensures that our estimates reflect the effects of disclosure itself rather than differences in the underlying AI applications. Details on the extraction procedure are provided in Section 4.4 and Online Appendix B3. c) Robust standard errors clustered at the category level. * $p<0.1$, ** $p<0.05$, *** $p<0.01$.

Table 3 summarizes the moderating effects of AI disclosure strategies on crowdfunding outcomes. For the substantive signal, the coefficient of the interaction term *AIDisclosure × HighAIInvolvement* is significantly negative in Columns 1 and 5, indicating that greater AI involvement amplifies the negative effect of AI disclosure on funding performance. This finding suggests that backers respond more favorably when AI is mainly used to assist, rather than replace, human effort.

Turning to rhetorical signals, we observe three key findings. First, higher levels of explicitness in AI disclosures are associated with improved crowdfunding performance (Columns 2 and 5). When creators clearly articulate how AI is integrated into the project, they reduce uncertainty and enable potential backers to better assess project feasibility and creator commitment. Second, authenticity in disclosure contributes positively to funding outcomes (Columns 3 and 5). This suggests that, language that conveys sincerity and internal consistency tends to increase backer willingness to support the campaign. Third, and somewhat unexpectedly, greater positive emotion in AI disclosure is associated with lower crowdfunding performance (Columns 4 and 5). Although positive emotional framing is typically linked to favorable outcomes in other disclosure contexts, such as financial reporting (Feldman et al., 2010), prosocial behaviors (Moran & Bagchi, 2019), and livestreaming shopping (Lin et al., 2021), our finding deviates from this pattern. One possible explanation is that, given the high complexity and opacity of AI technologies, overly enthusiastic descriptions of AI usage and potential may lead backers to suspect "AI washing",[12] where exaggerated claims about AI capabilities create doubts about the credibility and prospects of AI-involved projects.

---

[12] For discussions of AI washing, see more on https://www.techtarget.com/whatis/feature/AI-washing-explained-Everything-you-need-to-know (accessed June 5, 2025).

On the whole, while AI disclosure negatively impacts campaign success and total funding raised, strategic framing can mitigate these effects, highlighting the role of substantive and rhetorical signals in shaping backers' perceptions. Together, our findings extend signaling theory by showing how AI-related transparency shapes backers' responses and how strategic framing can help counter the downsides of disclosure.

**6. MECHANISM EXPLORATION THROUGH SCENARIO-BASED EXPERIMENTS**

While previous analyses show that AI disclosure reduces crowdfunding performance, the mechanisms underlying this effect remain unclear. Drawing on prior research on signaling in crowdfunding (Blanchard et al., 2023; Kim et al., 2022; Steigenberger & Wilhelm, 2018) and work on AI-related social perception (Reif et al., 2025), we propose that AI disclosure may change how backers evaluate both the project and its creator. Accordingly, we focus on two specific mechanisms: perceived AI washing and perceived creator competence. First, backers may interpret AI disclosure as an attempt to artificially inflate the technological sophistication of a campaign, a phenomenon we refer to as *AI washing*. Such perceptions can raise concerns about authenticity or strategic obfuscation, thereby undermining backers' project evaluation. These concerns are particularly important in crowdfunding settings, where backers rely heavily on limited, informal, and often ambiguous signals to guide their decisions (Blanchard et al., 2023; Kim et al., 2022; Steigenberger & Wilhelm, 2018). Second, because AI reduces the effort required to complete tasks, prospective backers may infer that creators who rely on AI are less capable, less skillful, or possess fewer relevant resources. This interpretation aligns with recent evidence showing that AI adopters are often judged more harshly or perceived as less competent than those using traditional methods (Reif et al., 2025).

## 6.1. Experiment Design and Data Collection

To empirically examine these proposed mechanisms, we conduct a series of pre-registered,[13] scenario-based online experiments on Prolific (https://www.prolific.com/). These experiments are designed to causally identify how AI disclosure shapes backers' perceptions of both the project and its creator under varying moderator conditions. Specifically, we assess whether AI disclosure elevates concerns regarding AI washing and decreases perceptions of creator competence, two processes that may account for the negative performance effects observed in our main analyses.

We implement four separate experiments, each focused on one moderator: AI involvement, explicitness, authenticity, or positive emotion. Each experiment employs a between-subjects design in which participants are randomly assigned to either a high or low level of the focal moderator. All experimental stimuli are adapted from *actual* Kickstarter projects that feature AI disclosures. To ensure precise manipulation of the focal dimension, we refined the disclosure language using ChatGPT to generate a counterfactual version: when the original project text reflected a high level of the moderator, we rephrased it to produce a low-level counterpart (and vice versa), while minimizing changes to other textual features.

Each experiment follows a consistent three-step procedure. *Step 1*: Participants review brief study instructions. *Step 2*: They read a crowdfunding project description that includes an AI disclosure tailored to their assigned condition. The disclosure wording is systematically varied to reflect the intended level of involvement, explicitness, authenticity, or emotional tone. We administer a manipulation check to verify that the perceived levels of the moderator align with the assigned condition. *Step 3*: Participants report their pledge intention, which serves as our primary dependent variable, along with their perceptions of the project and its creator, including perceived AI washing (Haridasan et al. 2015) and perceived creator competence (Kuan & Chau, 2001), all measured on 7-point Likert scales. This structured, three-step design allows us to isolate how specific features of AI disclosure influence backers' psychological responses (the proposed mediators) and their support intentions. Additional procedural details and full experimental materials are provided in Online Appendix E.

---

[13] All hypotheses, materials, and analysis plans were pre-registered.

### 6.2. Manipulation Check

To ensure data quality and reduce linguistic or cultural confounds, we applied stringent prescreening criteria on Prolific. Participants were required to have a prior approval rate of 95–100%, over 50 previous submissions, English as their first language, and both country of birth and nationality restricted to the United States. These criteria help ensure that respondents can reliably interpret disclosure language and that any observed effects are not attributable to differences in comprehension or cultural background. Each participant was permitted to take part in only one of the four experiments to avoid potential cross-exposure or contamination across studies.

We conducted four experiments, each manipulating one disclosure signal at two levels (high vs. low). In total, we obtained complete responses from 318 participants across the four experiments, with approximately 40 participants assigned to each condition. A randomization check (Online Appendix F) confirms that demographic and prior crowdfunding experience variables are balanced across conditions, supporting the validity of our randomization process. Among these respondents, 52 failed the manipulation check and were excluded from further analysis. Our subsequent regression results are based on the remaining 266 participants.[14]

---

[14] As a robustness check, we conducted regression analyses using the full sample, including participants who failed the attention check. Results remain highly consistent.

Table 4 presents the results of our manipulation checks. Participants evaluated the construct associated with the experiment to which they were assigned, indicating whether the focal attribute appeared at a high or low level based on their perceptions. Responses to the manipulation check question were coded as 1 for high-level perceived values and 0 for low-level perceived values of the construct. Across all four experiments, the manipulation checks were successful.

**Table 4. Manipulation Check**

| Construct | Sample size | | Mean (SD) | | $t$-statistic |
|---|---|---|---|---|---|
| | High condition | Low condition | High condition | Low condition | |
| AI Involvement | 41 | 38 | 0.88 (0.33) | 0.13 (0.34) | 9.84*** |
| Explicitness | 42 | 37 | 0.83 (0.38) | 0.14 (0.35) | 8.53*** |
| Authenticity | 41 | 39 | 0.88 (0.33) | 0.23 (0.43) | 7.60*** |
| Positive Emotion | 38 | 42 | 0.87 (0.34) | 0.26 (0.45) | 6.78*** |

Note: * $p<0.1$, ** $p<0.05$, *** $p<0.01$

For AI involvement, participants in the high condition were significantly more likely to report a high level of AI involvement (M = 0.88, SD = 0.33) than those in the low condition (M = 0.13, SD = 0.34; $t$ = 9.84, $p$ < 0.01), with 87.8 percent and 86.8 percent correctly identifying the intended levels, respectively. For explicitness, the high condition was again viewed as more explicit (M = 0.83, SD = 0.38) than the low condition (M = 0.14, SD = 0.35; $t$ = 8.53, $p$ < 0.01), and 83.3 percent and 86.5 percent of participants correctly identified the intended levels. For authenticity, participants in the high condition similarly rated disclosures as significantly more authentic (M = 0.88, SD = 0.33) than those in the low condition (M = 0.23, SD = 0.43; $t$ = 7.60, $p$ < 0.01), with corresponding correct identification rates of 87.8 percent and 76.9 percent. Finally, for emotional tone, the high condition elicited significantly greater perceived excessive positive emotionality (M = 0.87, SD = 0.34) than the low condition (M = 0.26, SD = 0.45; $t$ = 6.78, $p$ < 0.01), with correct identification rates of 86.8 percent and 73.8 percent, respectively. Overall, these results confirm that participants consistently detected the intended high versus low manipulations across all constructs.

### 6.3. Experimental Results

The following results are based on the 266 participants who passed the manipulation check, including 69 in the AI involvement module, 67 in the explicitness module, 66 in the authenticity module, and 64 in the emotion module.

**6.3.1. Results on AI Involvement**

As illustrated in Figure 3, projects involving higher levels of AI elicited noticeably different reactions from participants. Those evaluating high-AI-involvement projects are far less willing to pledge than those reviewing projects with limited AI involvement (M = 1.9 vs. 4.6, $p < 0.01$). They also judge the creator as significantly less competent (M = 2.6 vs. 5.8, $p < 0.01$), while the increase in perceived AI washing, though directionally consistent, is relatively modest and insignificant (M = 3.0 vs. 2.5, n.s.).

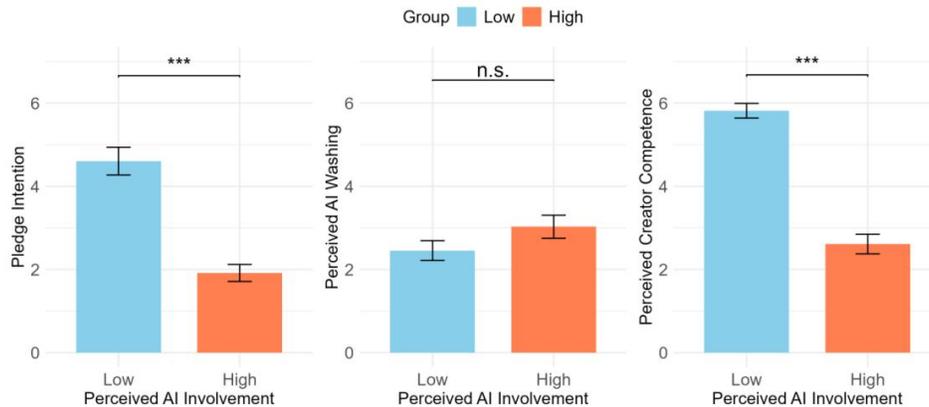

**Figure 3. Effects of Low vs. High AI Involvement**

We further test the potential underlying mechanism with a sequential mediation analysis using AI involvement as the independent variable, pledge intention as the dependent variable, and perceived AI washing and perceived creator competence as serial mediators (Model 6; Hayes, 2017). As shown in Figure 4, high AI involvement produces only a slight rise in AI washing (b = 0.57, n.s.), and this shift has little bearing on perceived creator competence (b = −0.06, n.s.). By contrast, high AI involvement has a strong and negative direct effect on competence (b = −3.17, $p < 0.01$), which in turn is a key driver of decreased pledge intention (b = 0.71, $p < 0.01$). Neither the direct pathway from high AI involvement to pledge intention (b = −0.36, n.s.) nor the link from AI washing to pledge intention (b = −0.08, n.s.) contributes meaningfully. Based on the effect sizes summarized in Table G1 of Online Appendix G, the indirect effect of high AI involvement on pledge intention through perceived creator competence is −2.263 (bootstrapped 95% CI: [-3.402, -1.314]), while the total effect of high AI involvement on pledge intention is −2.689 ($p < 0.01$), indicating that diminished perceptions of creator competence account for the majority of the negative impact of high AI involvement on backers' support.

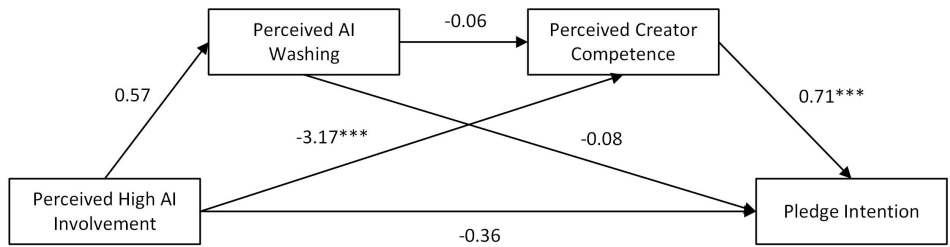

**Figure 4. Treatment Effect of AI Involvement on Pledge Intention, Mediated by AI Washing and Creator Competence**

### 6.3.2. Results on Explicitness

Figure 5 shows that high explicitness in AI disclosure produces noticeably more favorable evaluations. Participants evaluating projects with high-explicitness AI disclosures are more willing to pledge than those exposed to low-explicitness statements (M = 4.8 vs. 3.2, $p < 0.01$). These participants also form more positive impressions of the creator, reporting lower perceptions of AI washing (M = 2.1 vs. 2.9, $p < 0.05$) and higher perceived creator competence (M = 5.8 vs. 3.8, $p < 0.01$).

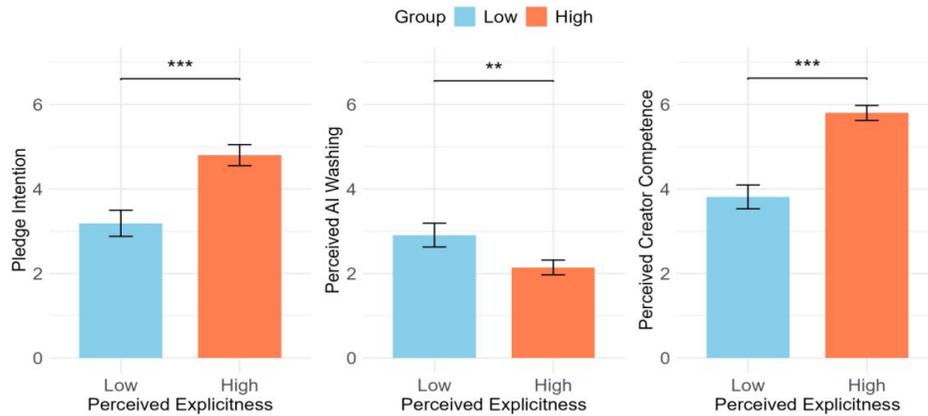

**Figure 5. Effects of Low vs. High Perceived Explicitness**

Evidence from Figure 6 further clarifies how AI disclosure explicitness shapes these outcomes. Higher explicitness directly increases perceived creator competence (b = 1.61, $p <$ 0.01), which in turn elevates willingness to pledge (b = 0.76, $p <$ 0.01). Additionally, higher explicitness also reduces perceptions of AI washing (b = −0.76, $p <$ 0.05), and lower AI washing is tied to higher perceived competence (b = −0.49, $p <$ 0.01).

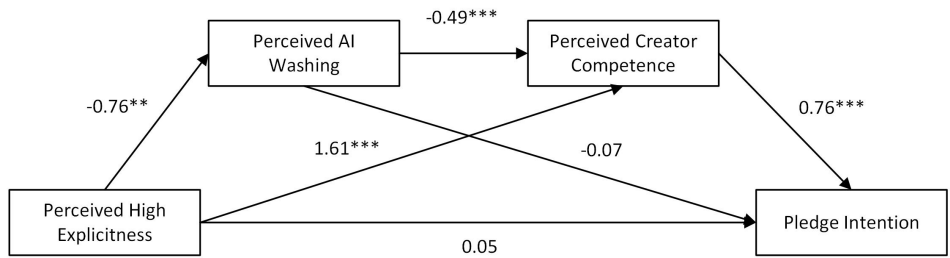

**Figure 6. Treatment Effect of Perceived Explicitness on Pledge Intention, Mediated by Perceived AI Washing and Perceived Creator Competence**

By contrast, the direct pathway from explicitness to pledge intention (b = 0.05, n.s.) and the link from AI washing to pledge intention (b = −0.07, n.s.) are not significant. This pattern indicates that explicitness strengthens support primarily by enhancing perceived competence.

Table G2 of Online Appendix G further clarifies how AI disclosure explicitness shapes these outcomes. High explicitness in the AI disclosure primarily increases pledge intention by enhancing backers' perceptions of creator competence (indirect effect = 1.226, bootstrapped 95%

CI: [0.584, 1.991]), and the total effect is 1.612 ($p < 0.01$). Meanwhile, high explicitness also exerts an additional positive effect on pledge intention through a sequential mediation pathway (Perceived High Explicitness → Perceived AI Washing → Perceived Creator Competence → Pledge Intention; indirect effect = 0.286, bootstrapped 95% CI: [0.042, 0.590]).

### 6.3.3. Results on Authenticity

Figure 7 presents the effects of perceived authenticity on pledge intention and perceived creator competence. Participants exposed to high-authenticity disclosures express significantly greater willingness to pledge than those exposed to low-authenticity disclosures (M = 4.3 vs. 2.7, $p < 0.01$). They are also less likely to perceive AI washing (M = 2.4 vs. 3.8, $p < 0.01$) and judge the creator to be more competent (M = 5.1 vs. 3.7, $p < 0.01$).

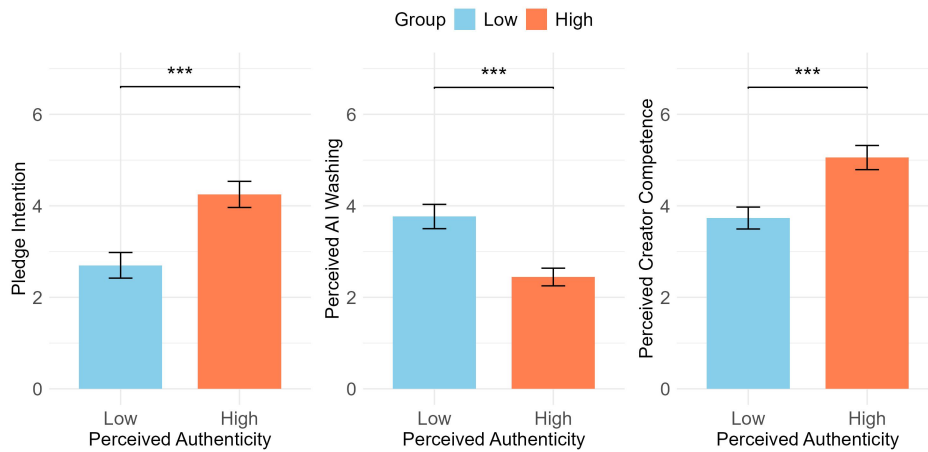

**Figure 7. Effects of Low vs. High Perceived Authenticity**

The mediation analysis results presented in Figure 8 further illustrate how authenticity shapes these outcomes. High authenticity in the AI disclosure statement strongly reduces perceptions of AI washing (b = −1.32, $p < 0.01$), and lower perceived AI washing is associated with higher perceived competence (b = −0.67, $p < 0.01$). High authenticity also directly improves perceived creator competence (b = 0.86, $p < 0.01$).

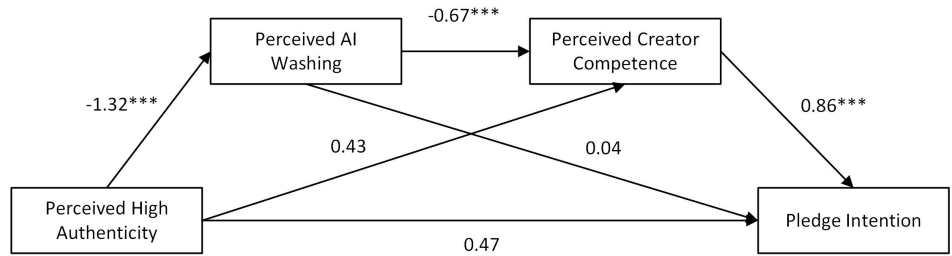

**Figure 8. Treatment Effect of Perceived Authenticity on Pledge Intention, Mediated by Perceived AI Washing and Perceived Creator Competence**

As summarized in Table G3 of Online Appendix G, neither the direct effect of high authenticity on pledge intention (b = 0.469, n.s.) nor simple mediation through perceived competence (indirect effect=0.371, bootstrapped 95% CI: [-0.244, 0.953]) or AI washing (indirect effect = -0.051, bootstrapped 95% CI: [ -0.479, 0.339]) alone proves significant. Instead, high authenticity operates through sequential mediation: reducing AI washing perceptions, which enhances competence perceptions and increases pledge intention (Perceived High Authenticity → Perceived AI Washing → Perceived Creator Competence → Pledge Intention; indirect effect = 0.761, bootstrapped 95% CI: [0.335, 1.351]).

### 6.3.4. Results on Emotional Tone

Figure 9 shows that excessively positive emotional tone leads to markedly less favorable reactions from participants. Individuals exposed to AI disclosure statements containing highly positive emotion are significantly less willing to pledge than those encountering more moderately worded statements (M = 3.2 vs. 4.6, $p < 0.01$). They also form more negative impressions of the project and creator, reporting higher perceptions of AI washing (M = 5.5 vs. 2.3, $p < 0.01$) and lower perceived creator competence (M = 3.6 vs. 5.4, $p < 0.01$).

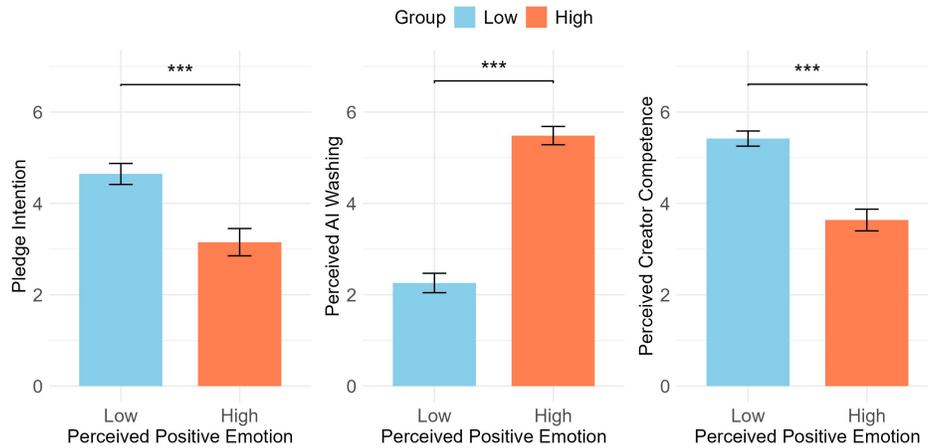

**Figure 9. Effects of Low vs. High Positive Emotional Tone**

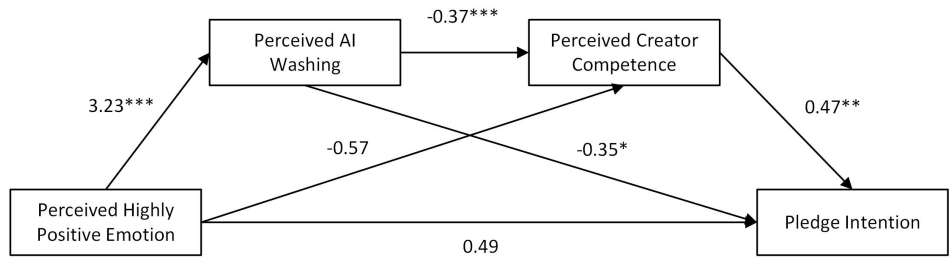

**Figure 10. Treatment Effect of Perceived Highly Positive Emotion on Pledge Intention as Mediated by Perceived AI Washing and Perceived Creator Competence**

Figure 10 and Table G4 further illustrate why excessive positive emotion in the AI disclosure statement can produce unintended negative effects on backers' pledge intention. In particular, excessively positive emotion does not directly affect pledge intention or perceived creator competence; instead, it exerts a significant indirect effect through increased AI washing perceptions and the resulting decrease in perceived creator competence (Perceived Highly Positive Emotion → Perceived AI Washing → Perceived Creator Competence → Pledge Intention). This sequential indirect effect is sizable (b = −0.572, bootstrapped 95% CI: [-1.558, -0.054]). Thus, rather than attracting backer engagement, overly positive emotional framing can backfire by triggering AI washing concerns that undermine competence perceptions.

# 7. ROBUSTNESS CHECKS

To verify the robustness of our results, we conduct several additional checks, summarized in Table 5. Specifically, we re-estimate the models using a two-stage propensity score matching (PSM; Rosenbaum & Rubin, 1983) approach, rerun the analysis using campaign launch date instead of end date to define treatment status, and examine whether the findings hold with alternative dependent variables and additional controls. We further assess the robustness of the emotional tone results using two alternative approaches: (1) applying an alternative sentiment analysis model (SieBERT) and (2) re-performing the text classifications using Claude-Sonnet-4, an advanced LLM developed by Anthropic. The results remain highly consistent across all specifications, lending strong support to our main findings.

**Table 5. Summary of Robustness Checks**

| Analysis | Objective | Location |
|---|---|---|
| Alternative DID setup | Employing a two-stage PSM approach to construct an alternative | 7.1 |
| Alternative timing definition | Using the campaign launch date to classify the treatment period | 7.2 |
| Alternative dependent variables | Robustness of results to alternative dependent variable | 7.3 |
| Additional control variables | Robustness of results to additional control variables | 7.4 |
| Alternative method for sentiment detection | Using SieBERT to assess the emotional tone of AI disclosure statements | 7.5 |
| Alternative LLM for Labeling | Robustness check using labels generated by Claude-Sonnet-4 | 7.6 |

## 7.1. Alternative DID Setup

In our main analysis, we adopt a traditional keyword-based approach to classify projects into treatment and control groups. This method allows for a straightforward and transparent way to identify AI-related projects. However, it identifies whether projects mention AI without capturing the deployment or substantive use of AI. As a robustness check, we employ a two-stage PSM procedure to identify projects similar to those that disclose AI usage and use them as counterfactuals.

Figure 11 illustrates the matching process and sample construction. In the first stage, we use one-to-one PSM to match projects posted before the disclosure policy implementation with those that include AI disclosure statements in the post-policy period; these pre-policy matches serve as the pre-policy counterparts for the AI-disclosed projects. In the second stage, we construct the control group by selecting projects that do not disclose AI usage and applying one-to-one PSM again to identify their corresponding pre-policy counterparts (Liang et al., 2025). This two-stage procedure ensures that both the treatment and control groups are paired with comparable pre-policy projects, thereby enabling a credible counterfactual comparison.

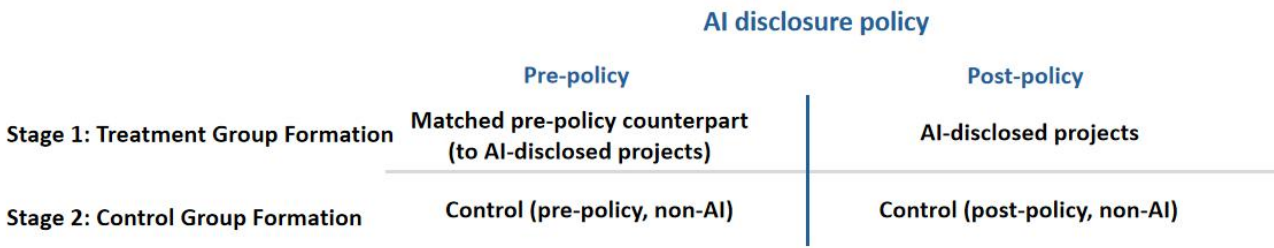

**Figure 11. Two-stage PSM**

It should be noted that our approach relies on the assumption that creators truthfully disclose their use of AI in the post-policy period. This assumption is supported by Kickstarter's enforcement practices: the platform may suspend projects that fail to disclose AI usage, and creators who misrepresent or attempt to bypass the disclosure requirement may be prohibited from submitting future projects.[15] Kickstarter also employs human reviewers who can request additional evidence when a project's AI usage is unclear, rather than relying solely on creators' self-reported statements. Importantly, any remaining non-compliance (e.g., creators who use AI but do not disclose it) would bias our estimates toward zero, making our treatment effect more conservative.

---

[15] As stated publicly by Kickstarter, "If creators don't properly disclose their use of AI during the submission process, Kickstarter may suspend the project. Those who try to bypass Kickstarter's policies or purposefully misrepresent their project won't be allowed to submit." See more on https://www.engadget.com/kickstarter-projects-will-soon-have-to-disclose-any-ai-use-145100394.html and https://www.neowin.net/news/kickstarter-to-require-ai-transparency-from-creatives/ (accessed November 17, 2025).

Our final sample comprises 4,880 projects, including 2,440 treatment projects and 2,440 matched control projects. Figure 12 presents the parallel trends test using the two-stage PSM sample.[16] Our results show that the treatment and control groups followed parallel trends prior to the policy implementation and exhibited comparable crowdfunding performance during the pre-policy period. Following the policy mandate, however, projects with AI disclosure experienced significantly more negative crowdfunding outcomes than their matched counterparts, consistent with our main findings.

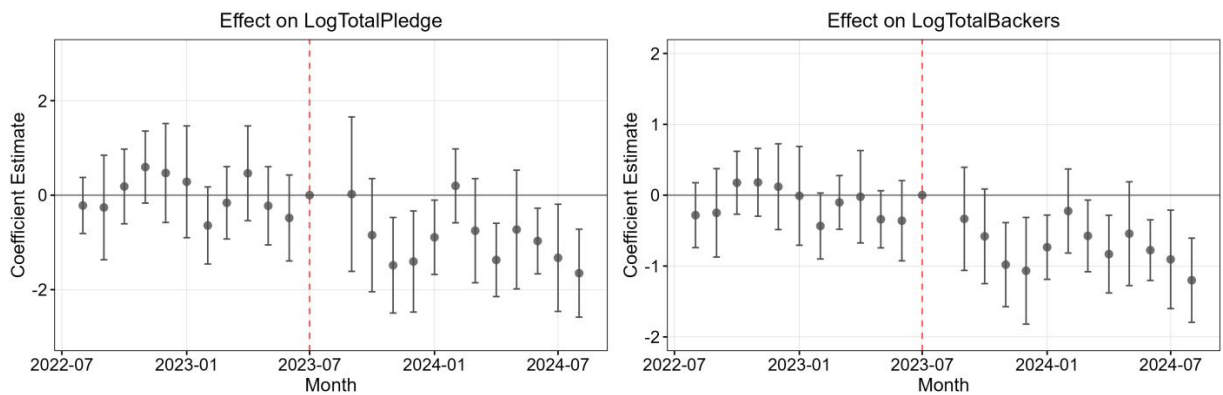

**Figure 12. Parallel Trend Test Based on a Two-Stage PSM Sample**

Table 6 reports the average treatment effect of AI disclosure on crowdfunding success using LLM-classified AI involvement. Table 7 reports heterogeneous treatment effects, which are highly consistent with our main findings. These results confirm that our findings are robust to alternative measurement approaches.

**Table 6. Main Effect Based on a Two-Stage PSM Sample**

| | *Dependent variable:* |
|---|---|

---

[16] No treated project ends in August 2023 because the policy became effective on August 29, 2023, leaving only a few days in that month for projects to close under the new disclosure requirement. Since crowdfunding campaigns typically span several weeks, almost all projects affected by the mandate closed in September 2023 or later. This timing explains the absence of treated projects with an end date in August 2023 and is consistent with the implementation window of the policy shock.

|  | LogTotalPledge | LogTotalBackers |
|---|---|---|
|  | (1) | (2) |
| Treatment | -0.064 | -0.042 |
|  | (0.105) | (0.054) |
| After | -0.349 | -0.157 |
|  | (0.341) | (0.184) |
| Treatment × After | -1.003*** | -0.643*** |
|  | (0.200) | (0.123) |
| Control Variables | YES | YES |
| Category FE | YES | YES |
| Launch Month FE | YES | YES |
| Day-of-week FE | YES | YES |
| Observations | 4,880 | 4,880 |
| R-squared | 0.299 | 0.371 |

Notes: a) We include the same project and creator controls as Table 2. Coefficients are omitted for brevity. b) See Table 2 for the *After* dummy definition. c) Robust standard errors clustered at the category level. * $p<0.1$, ** $p<0.05$, *** $p<0.01$.

**Table 7. Heterogeneous Treatment Effect Based on a Two-Stage PSM Sample (with *LogTotalPledge* as Dependent Variable)**

|  | *Dependent variable:* | | | | |
|---|---|---|---|---|---|
|  | LogTotalPledge | | | | |
|  | (1) | (2) | (3) | (4) | (5) |
| AIDisclosure | -1.213*** | -1.625*** | -1.759*** | -1.084*** | -1.564*** |
|  | (0.164) | (0.192) | (0.156) | (0.161) | (0.152) |
| AIDisclosure × HighAIInvolvement | -0.592** |  |  |  | -0.660*** |
|  | (0.244) |  |  |  | (0.198) |
| AIDisclosure × HighExplicitness |  | 0.846*** |  |  | 0.716** |
|  |  | (0.209) |  |  | (0.300) |
| AIDisclosure × HighAuthenticity |  |  | 0.987*** |  | 0.579* |
|  |  |  | (0.219) |  | (0.272) |
| AIDisclosure × HighPosEmotion |  |  |  | -0.505*** | -0.446*** |
|  |  |  |  | (0.125) | (0.149) |
| Control Variables | YES | YES | YES | YES | YES |
| AI Topic Controls | YES | YES | YES | YES | YES |
| Category FE | YES | YES | YES | YES | YES |
| Launch Month FE | YES | YES | YES | YES | YES |
| Day-of-week FE | YES | YES | YES | YES | YES |
| Observations | 2,580 | 2,580 | 2,580 | 2,580 | 2,580 |
| R-squared | 0.327 | 0.332 | 0.331 | 0.326 | 0.341 |

Notes: a) We include the same project and creator controls as Table 2. Coefficients are omitted for brevity. b) We include the same AI topic controls as Table 3. Related variable definitions and coefficients are omitted for brevity. c) For brevity, we report only the results with *LogTotalPledge* as the dependent variable; results with *LogTotalBackers* as the dependent variable are highly consistent and reported in Online Appendix H. d) Robust standard errors clustered at the category level. * $p<0.1$, ** $p<0.05$, *** $p<0.01$.

### 7.2. Alternative Timing Definition

In our main analysis, we rely on each project's campaign end date to determine whether a project belongs to the pre- or post-policy period, ensuring a clear model setup and precise identification of policy exposure. To further assess robustness, we use the campaign launch time as the temporal reference point. Specifically, we include all projects launched between August 1, 2022, and August 31, 2024, as our alternative sample, excluding those launched before but closing after the policy introduction. This procedure yields a final sample of 33,583 projects, comprising 5,138 treatment projects and 28,445 control projects. As shown in Tables 8 and 9, the results based on this alternative specification remain highly consistent with our main findings.

**Table 8. Main Effect Based on Alternative Time Variable**

|  | Dependent variable: | |
|---|---|---|
|  | LogTotalPledge | LogTotalBackers |
|  | (1) | (2) |
| Treatment | 0.210*** | 0.069* |
|  | (0.067) | (0.033) |
| Treatment × After | -0.500*** | -0.276*** |
|  | (0.094) | (0.041) |
| Control Variables | YES | YES |
| Category FE | YES | YES |
| Launch Month FE | YES | YES |
| Day-of-week FE | YES | YES |
| Observations | 33,583 | 33,583 |
| R-squared | 0.249 | 0.305 |

Notes: a) We include the same project and creator controls as Table 2. Coefficients are omitted for brevity. b) See Table 2 for the *After* dummy definition. c) Robust standard errors clustered at the category level. * $p<0.1$, ** $p<0.05$, *** $p<0.01$.

**Table 9. Heterogeneous Treatment Effect Based on an Alternative Time Variable (with *LogTotalPledge* as Dependent Variable)**

|  | Dependent variable: | | | | |
|---|---|---|---|---|---|
|  | LogTotalPledge | | | | |
|  | (1) | (2) | (3) | (4) | (5) |
| AIDisclosure | -1.374*** | -1.883*** | -2.015*** | -1.254*** | -1.763*** |
|  | (0.201) | (0.237) | (0.120) | (0.167) | (0.117) |
| AIDisclosure × HighAIInvolvement | -0.746** |  |  |  | -0.745*** |
|  | (0.261) |  |  |  | (0.197) |
| AIDisclosure × HighExplicitness |  | 0.992*** |  |  | 0.761** |
|  |  | (0.194) |  |  | (0.303) |
| AIDisclosure × HighAuthenticity |  |  | 1.244*** |  | 0.775*** |
|  |  |  | (0.195) |  | (0.261) |
| AIDisclosure × HighPosEmotion |  |  |  | -0.589*** | -0.489*** |
|  |  |  |  | (0.132) | (0.150) |
| Control Variables | YES | YES | YES | YES | YES |
| AI Topic Controls | YES | YES | YES | YES | YES |
| Category FE | YES | YES | YES | YES | YES |
| Launch Month FE | YES | YES | YES | YES | YES |
| Day-of-week FE | YES | YES | YES | YES | YES |
| Observations | 19,360 | 19,360 | 19,360 | 19,360 | 19,360 |
| R-squared | 0.269 | 0.271 | 0.270 | 0.269 | 0.273 |

Notes: a) We include the same project and creator controls as Table 2. Coefficients are omitted for brevity. b) We include the same AI topic controls as Table 3. Related variable definitions and coefficients are omitted for brevity. c) For brevity, we report only the results with *LogTotalPledge* as the dependent variable; results with *LogTotalBackers* as the dependent variable are highly consistent and reported in Online Appendix H. d) Robust standard errors clustered at the category level. * $p<0.1$, ** $p<0.05$, *** $p<0.01$.

### 7.3. Alternative Dependent Variables

In the main analysis, we use the total amount raised and the total number of backers as primary dependent variables. Given Kickstarter's all-or-nothing funding model, where entrepreneurs receive funds only if they reach their predetermined funding goal within the campaign duration (Calic et al., 2023), we examine two additional outcome variables: *CampaignSuccess* and *LogpledgedAdj*. *CampaignSuccess* is a binary indicator of whether a project ultimately met its funding goal, and *LogpledgedAdj* captures the realized funding amount under the all-or-nothing policy (Soublière & Gehman, 2020). As shown in Tables 10 and 11, the results based on this alternative outcome measure remain consistent with our main findings.

**Table 10. Main Effect Based on Alternative Dependent Variables**

|  | *Dependent variable:* | |
|---|---|---|
|  | CampaignSuccess | LogPledgedAdj |
| Treatment | 0.011 | 0.216** |
|  | (0.010) | (0.096) |
| After | -0.057* | -0.529* |
|  | (0.030) | (0.263) |
| Treatment × After | -0.057*** | -0.634*** |
|  | (0.007) | (0.080) |
| Control Variables | YES | YES |
| Category FE | YES | YES |
| Launch Month FE | YES | YES |
| Day-of-week FE | YES | YES |
| Observations | 35,832 | 35,832 |
| R-squared | 0.309 | 0.264 |

Notes: a) We include the same project and creator controls as Table 2. Coefficients are omitted for brevity. b) See Table 2 for the *After* dummy definition. c) Robust standard errors clustered at the category level. * $p<0.1$, ** $p<0.05$, *** $p<0.01$.

**Table 11. Heterogeneous Treatment Effects Based on Alternative Dependent Variables**

|  | *Dependent variable:* | | | | |
|---|---|---|---|---|---|
|  | CampaignSuccess | | | | |
|  | (1) | (2) | (3) | (4) | (5) |
| AIDisclosure | -0.174*** | -0.221*** | -0.243*** | -0.162*** | -0.209*** |
|  | (0.019) | (0.017) | (0.024) | (0.017) | (0.021) |
| AIDisclosure × HighAIInvolvement | -0.093** |  |  |  | -0.086*** |

|  | (0.034) |  |  |  | (0.029) |
| --- | --- | --- | --- | --- | --- |
| AIDisclosure × HighExplicitness |  | 0.081*** |  |  | 0.055** |
|  |  | (0.016) |  |  | (0.020) |
| AIDisclosure × HighAuthenticity |  |  | 0.128*** |  | 0.088*** |
|  |  |  | (0.022) |  | (0.024) |
| AIDisclosure × HighPosEmotion |  |  |  | -0.065*** | -0.053** |
|  |  |  |  | (0.020) | (0.021) |
| Control Variables | YES | YES | YES | YES | YES |
| AI Topic Controls | YES | YES | YES | YES | YES |
| Category FE | YES | YES | YES | YES | YES |
| Launch Month FE | YES | YES | YES | YES | YES |
| Day-of-week FE | YES | YES | YES | YES | YES |
| Observations | 20,514 | 20,514 | 20,514 | 20,514 | 20,514 |
| R-squared | 0.329 | 0.329 | 0.330 | 0.329 | 0.330 |

Table 11. (continued)

|  | Dependent variable: LogPledgedAdj | | | | |
| --- | --- | --- | --- | --- | --- |
|  | (1) | (2) | (3) | (4) | (5) |
| AIDisclosure | -1.879*** | -2.431*** | -2.647*** | -1.745*** | -2.258*** |
|  | (0.171) | (0.182) | (0.188) | (0.130) | (0.154) |
| AIDisclosure × HighAIInvolvement | -1.056*** |  |  |  | -1.001*** |
|  | (0.354) |  |  |  | (0.310) |
| AIDisclosure × HighExplicitness |  | 0.948*** |  |  | 0.677** |
|  |  | (0.210) |  |  | (0.242) |
| AIDisclosure × HighAuthenticity |  |  | 1.409*** |  | 0.934*** |
|  |  |  | (0.229) |  | (0.203) |
| AIDisclosure × HighPosEmotion |  |  |  | -0.754*** | -0.625** |
|  |  |  |  | (0.199) | (0.217) |
| Control Variables | YES | YES | YES | YES | YES |
| AI Topic Controls | YES | YES | YES | YES | YES |
| Category FE | YES | YES | YES | YES | YES |
| Launch Month FE | YES | YES | YES | YES | YES |
| Day-of-week FE | YES | YES | YES | YES | YES |
| Observations | 20,514 | 20,514 | 20,514 | 20,514 | 20,514 |
| R-squared | 0.285 | 0.285 | 0.286 | 0.285 | 0.287 |

Notes: a) We include the same project and creator controls as Table 2. Coefficients are omitted for brevity. b) We include the same AI topic controls as Table 3. Related variable definitions and coefficients are omitted for brevity. c) Robust standard errors clustered at the category level. * $p<0.1$, ** $p<0.05$, *** $p<0.01$.

### 7.4. Additional Control Variables

To account for rhetorical variation in project descriptions beyond the AI disclosure section, we apply the same coding framework to extract style-related textual features from the full project narrative.[17] These measures are included as additional controls to isolate the effects of AI disclosure from general stylistic differences across campaigns. As shown in Tables 12 and 13, after accounting for these factors, our results remain qualitatively unchanged.

---

[17] Following the same approach used to extract text features in the AI disclosure, we apply the same GPT-4o-mini prompt to measure the description's explicitness and authenticity, and use VADER to quantify its emotional tone.

**Table 12. Main Effect with Additional Control Variables**

|  | Dependent variable: | |
|---|---|---|
|  | LogTotalPledge | LogTotalBackers |
|  | (1) | (2) |
| Treatment | 0.195*** | 0.062 |
|  | (0.053) | (0.035) |
| After | -0.443** | -0.272*** |
|  | (0.161) | (0.081) |
| Treatment × After | -0.472*** | -0.252*** |
|  | (0.088) | (0.040) |
| Control Variables | YES | YES |
| Additional Project Description Controls | YES | YES |
| Category FE | YES | YES |
| Launch Month FE | YES | YES |
| Day-of-week FE | YES | YES |
| Observations | 35,832 | 35,832 |
| R-squared | 0.265 | 0.320 |

Notes: a) We include the same project and creator controls as Table 2. Coefficients are omitted for brevity. b) See Table 2 for the *After* dummy definition. c) To ensure that our estimates are not driven by stylistic variation in how project creators write their descriptions, we also extract linguistic features directly from the project descriptions. Using the same GPT-4o-mini prompts applied to the AI-disclosure statements, we obtain measures of explicitness and authenticity, and use VADER to compute sentiment scores. These variables are included as additional controls to separate the effect of disclosure style from general writing style. d) Robust standard errors clustered at the category level. * $p<0.1$, ** $p<0.05$, *** $p<0.01$.

**Table 13. Heterogeneous Treatment Effects with Additional Control Variables (with *LogTotalPledge* as Dependent Variable)**

|  | Dependent variable: | | | | |
|---|---|---|---|---|---|
|  | LogTotalPledge | | | | |
|  | (1) | (2) | (3) | (4) | (5) |
| AIDisclosure | -1.288*** | -1.795*** | -1.930*** | -1.181*** | -1.660*** |
|  | (0.178) | (0.221) | (0.129) | (0.149) | (0.119) |
| AIDisclosure × HighAIInvolvement | -0.822*** |  |  |  | -0.810*** |
|  | (0.254) |  |  |  | (0.197) |
| AIDisclosure × HighExplicitness |  | 0.941*** |  |  | 0.731** |
|  |  | (0.207) |  |  | (0.321) |
| AIDisclosure × HighAuthenticity |  |  | 1.203*** |  | 0.737** |
|  |  |  | (0.190) |  | (0.261) |
| AIDisclosure × HighPosEmotion |  |  |  | -0.592*** | -0.485*** |
|  |  |  |  | (0.144) | (0.162) |
| Control Variables | YES | YES | YES | YES | YES |
| AI Topic Controls | YES | YES | YES | YES | YES |
| Project Description Controls | YES | YES | YES | YES | YES |
| Category FE | YES | YES | YES | YES | YES |
| Launch Month FE | YES | YES | YES | YES | YES |
| Day-of-week FE | YES | YES | YES | YES | YES |
| Observations | 20,514 | 20,514 | 20,514 | 20,514 | 20,514 |
| R-squared | 0.281 | 0.282 | 0.283 | 0.281 | 0.284 |

Notes: a) We include the same project and creator controls as Table 2. Coefficients are omitted for brevity. b) We include the same AI topic controls as Table 3. Related variable definitions and coefficients are omitted for brevity. c) We include the same set of additional project description controls as described in Table 12. d) For brevity, we report only the results with *LogTotalPledge* as the dependent variable; results with *LogTotalBackers* as the dependent variable are reported in Appendix H. e) Robust standard errors clustered at the category level. * $p<0.1$, ** $p<0.05$, *** $p<0.01$.

## 7.5. Alternative Method for Sentiment Detection

In the main analysis, we use VADER, a widely adopted tool in the literature, to detect whether a project exhibits excessively positive sentiment in its AI disclosure statement based on the intensity of positive emotion. In this section, we employ SieBERT, an LLM fine-tuned for sentiment analysis (Brynjolfsson et al., 2025; Hartmann et al., 2023), to construct a binary indicator of whether the sentiment is positive. As shown in Table 14, we find that when using this alternative measure, the results remain consistent with our main findings.

**Table 14. Heterogeneous Treatment Effect Based on an Alternative Sentiment Tool (with *LogTotalPledge* as Dependent Variable)**

|  | Dependent variable: LogTotalPledge | | | | |
|---|---|---|---|---|---|
|  | (1) | (2) | (3) | (4) | (5) |
| AIDisclosure | -1.349*** | -1.864*** | -1.984*** | -1.073*** | -1.494*** |
|  | (0.183) | (0.225) | (0.129) | (0.303) | (0.228) |
| AIDisclosure × HighAIInvolvement | -0.794*** |  |  |  | -0.795*** |
|  | (0.255) |  |  |  | (0.208) |
| AIDisclosure × HighExplicitness |  | 0.979*** |  |  | 0.844*** |
|  |  | (0.206) |  |  | (0.284) |
| AIDisclosure × HighAuthenticity |  |  | 1.201*** |  | 0.715** |
|  |  |  | (0.197) |  | (0.261) |
| AIDisclosure × PosEmotion |  |  |  | -0.491** | -0.548** |
|  |  |  |  | (0.222) | (0.196) |
| Control Variables | YES | YES | YES | YES | YES |
| AI Topic Controls | YES | YES | YES | YES | YES |
| Category FE | YES | YES | YES | YES | YES |
| Launch Month FE | YES | YES | YES | YES | YES |
| Day-of-week FE | YES | YES | YES | YES | YES |
| Observations | 20,514 | 20,514 | 20,514 | 20,514 | 20,514 |
| R-squared | 0.265 | 0.266 | 0.267 | 0.264 | 0.268 |

Notes: a) Project and creator controls as described in Table 2. b) AI topic controls as described in Table 3. c) For brevity, we report only the results with *LogTotalPledge* as the dependent variable; results with *LogTotalBackers* as the dependent variable are highly consistent and reported in Appendix H. d) Robust standard errors clustered at the category level. * $p<0.1$, ** $p<0.05$, *** $p<0.01$.

## 7.6. Alternative LLM for Labeling

In the main analysis, we use GPT-4o-mini to label the levels of explicitness, authenticity, and AI involvement based on the AI disclosure statements. To examine whether our results are robust to the choice of LLM, we replicate the classification using Claude-Sonnet-4, a state-of-the-art model developed by Anthropic that emphasizes reasoning accuracy and reliability in text understanding[18]. As suggested by Table 15, the results based on labels generated by Claude-Sonnet-4 closely align with those obtained using GPT-4o-mini, indicating that our findings are stable across different LLM architectures and providers.

**Table 15. Heterogeneous Treatment Effect Based on an Alternative LLM for Labeling (with *LogTotalPledge* as Dependent Variable)**

|  | *Dependent variable:* | | | | |
|---|---|---|---|---|---|
|  | LogTotalPledge | | | | |
|  | (1) | (2) | (3) | (4) | (5) |
| AIDisclosure | -1.360*** | -2.248*** | -2.024*** | -1.227*** | -2.025*** |
|  | (0.192) | (0.138) | (0.164) | (0.149) | (0.172) |
| AIDisclosure × HighAIInvolvement | -1.037*** |  |  |  | -0.910*** |
|  | (0.272) |  |  |  | (0.180) |
| AIDisclosure × HighExplicitness |  | 1.474*** |  |  | 1.300*** |
|  |  | (0.172) |  |  | (0.140) |
| AIDisclosure × HighAuthenticity |  |  | 1.057*** |  | 0.307* |
|  |  |  | (0.239) |  | (0.168) |
| AIDisclosure × HighPosEmotion |  |  |  | -0.614*** | -0.398** |
|  |  |  |  | (0.140) | (0.160) |
| Control Variables | YES | YES | YES | YES | YES |
| AI Topic Controls | YES | YES | YES | YES | YES |
| Category FE | YES | YES | YES | YES | YES |
| Launch Month FE | YES | YES | YES | YES | YES |
| Day-of-week FE | YES | YES | YES | YES | YES |
| Observations | 20,514 | 20,514 | 20,514 | 20,514 | 20,514 |
| R-squared | 0.266 | 0.267 | 0.266 | 0.265 | 0.269 |

Notes: a) We include the same project and creator controls as Table 2. Coefficients are omitted for brevity. b) We include the same AI topic controls as Table 3. Related variable definitions and coefficients are omitted for brevity. c) For brevity, we report only the results with *LogTotalPledge* as the dependent variable; results with *LogTotalBackers* as the dependent variable are highly consistent and reported in Appendix H. d) Robust standard errors clustered at the category level. * $p<0.1$, ** $p<0.05$, *** $p<0.01$.

---

[18] Anthropic News is available at https://www.anthropic.com/news/claude-4 (accessed December 1,2025).

# 8. DISCUSSION AND CONCLUSION

## 8.1. Summary and Discussion

This study examines how platform-mandated AI disclosure affects crowdfunding outcomes and how its effects vary across different types of substantive and rhetorical signals. Using the policy shock as an exogenous source of variation, we identify the causal impact of AI disclosure and analyze the mechanisms through which it shapes backers' evaluations.

Our results show that disclosing AI use lowers the likelihood of project success, particularly when creators reveal high AI involvement. This finding suggests that transparency about extensive AI use may raise doubts about creator competence. In contrast, disclosing partial AI use appears to balance the appeal of innovation with the reassurance of human creation and oversight, resulting in weaker negative effects.

We also find that how creators frame AI disclosure matters greatly. Clear and specific statements (logos) and an authentic, credible tone (ethos) reduce investor skepticism and improve funding performance. However, disclosures that rely on highly positive or emotional language (pathos) often backfire, as they may be interpreted as exaggeration or hype. These results highlight that restrained, specific, factual communication is more effective than overly promotional framing when addressing backers' concerns about creator competence and potential deceptive AI positioning.

Our experimental mechanism analysis illuminates why these differential effects emerge. Substantive signals (AI involvement) operate primarily through competence perceptions: greater AI integration reduces perceived creator expertise. The three rhetorical signals, however, exhibit more complex causal pathways. (1) High explicitness in AI disclosure mainly bolsters pledge intention by directly enhancing perceived creator competence, while also having an indirect sequential mediation wherein explicit disclosure first reduces AI washing concerns, which subsequently elevates competence perceptions. (2) High authenticity influences pledge intentions predominantly via an indirect serial pathway: authentic communication alleviates AI washing concerns, which then elevates competence judgments. (3) Excessively positive emotional tone triggers the reverse serial process: it first heightens AI washing suspicions, which then erode competence perceptions. This reveals a critical asymmetry: promotional rhetoric intended to enhance perceptions actually backfires by activating suspicion that undermines the very judgments it aims to strengthen. Overall, our findings demonstrate that effective AI disclosure requires not merely transparency but strategic calibration of how information is framed. AI disclosure functions as a strategic communication tool that can build credibility when executed thoughtfully or undermine it when handled carelessly.

### 8.2. Managerial and Policy Implications

Our findings offer several implications for creators, platforms, and policymakers navigating the growing importance of AI transparency in digital markets. For creators, the results highlight that transparency should be managed strategically rather than mechanically. The key is to communicate AI involvement as a means of enhancing efficiency or creativity without undermining the human contribution. Disclosing limited or assistive AI use allows creators to signal both innovation and authenticity. Creators should avoid ambiguous or emotionally exaggerated claims, which may lead to perceptions of overstatement or opportunism. Instead, they should provide concrete, factual disclosure specifying AI functions, explaining their benefits, and emphasizing human oversight in ensuring quality and originality. Platforms could help by offering guidance or templates for effective disclosure statements, making it easier for creators to comply with policy while preserving trust.

For platforms, our findings emphasize the importance of balancing regulation with user experience. Platforms can play an active role in shaping how AI disclosure is perceived by introducing standardized formats, explanatory cues, or credibility indicators that clarify the meaning of disclosed AI use. For example, a visual tag system could distinguish between AI-assisted and AI-generated content, helping backers interpret disclosures accurately. Platforms could also use automated text screening to flag vague or misleading disclosures and prompt revisions before publication. Moreover, platforms may consider integrating creator education modules on AI communication ethics or authenticity, which could help sustain both creator success and overall marketplace integrity. Beyond compliance, platforms should recognize disclosure as part of the institutional architecture governing creator-backer relationships and sustained long-term platform engagement.

For policymakers, our evidence suggests that mandatory AI disclosure policies can have mixed effects. While such rules improve transparency and accountability, they may unintentionally discourage innovation if audiences interpret AI involvement as a negative quality signal. Policymakers should therefore complement disclosure requirements with initiatives that improve public literacy about AI, clarifying that partial or assistive use does not necessarily diminish creativity or reliability. A tiered or contextualized disclosure framework could also be useful, distinguishing different types and degrees of AI involvement rather than imposing a single, uniform requirement. This approach would help minimize overreaction to benign uses of AI and encourage truthful, nuanced reporting. Regulators might also collaborate with platforms to standardize definitions and best practices across industries, ensuring consistency and fairness (Ho et al. 2024) while maintaining flexibility for innovation.

Taken together, these implications underscore that AI disclosure is not only a compliance obligation but also a communication challenge. Creators and platforms should treat disclosure as a form of brand management and trust-building, where clarity, honesty, and proportionality are central. Policymakers, in turn, should design and enforce disclosure frameworks that encourage transparency without creating unnecessary fear or stigma toward AI adoption. Effective implementation across all three stakeholder groups can promote both accountability and innovation in AI-mediated digital ecosystems.

### 8.3. Limitations and Future Research

Our study has several limitations. First, our analysis is based on a reward-based crowdfunding platform for creative projects, where backers are motivated primarily by obtaining product rewards and supporting ventures they find compelling. The dynamics we observe may therefore differ in other contexts, such as donation-based platforms or social media content platforms. The salience of creator competence and AI washing concerns may vary across these contexts. Future research could compare these settings to assess whether the effects of AI disclosure generalize beyond creative campaigns and how contextual factors such as regulatory environments or cultural norms shape audience responses.

Moreover, our analysis focuses on short-term funding outcomes. Future work could examine longer-term consequences, such as creator reputation, repeat funding, or sustained community engagement. As AI disclosure becomes more common, researchers could also explore how creators calibrate their signaling strategies in response to evolving backer expectations, and how they integrate AI capabilities into project development.

*[Due to space limitations, the online appendices are hosted on the Open Science Framework (OSF) at https://osf.io/6bcva/overview?view_only=b378f0d57ec9412186292b41c8ea3d8c.]*